# Inflationary Models with Logarithmic Potentials

John D. Barrow and Paul Parsons

*Astronomy Centre, University of Sussex,
Brighton BN1 9QH, U. K.*

**Abstract**

We examine inflationary universe models driven by scalar fields with logarithmic potentials of the form $V(\phi) = V_0 \phi^p (\ln \phi)^q$. Combining the slow-roll approximation with asymptotic techniques, we identify regions of the potential where inflation may occur and obtain analytic expressions for the evolution of the field and the metric in these cases. We construct a family of exact solutions to the equations of motion with potentials resembling the above form and demonstrate their inflationary nature; deflationary and conflationary cosmological behaviours are also defined and identified. Finally, a computation of scalar density and tensor gravitational perturbations produced by the model is presented.

# 1 Introduction

The notion of scalar fields driving an epoch of accelerated expansion has become a widely accepted element of many early universe cosmologies [1]. The inflationary paradigm resolves many of the shortcomings of standard cosmological models whilst offering an explanation for the origin of structure in the cosmos [2] which is compelling and consistent with recent observations of the microwave background [3]. In the literature one may find many mutations of this idea, invoking wide-ranging physical conjectures to generate the negative stresses required for a period of inflation in the early universe. The simplest of these assumes the presence of a minimally coupled self-interacting scalar field $\phi$ with potential $V(\phi)$ evolving slowly in the presence of Einstein gravity [4 – 14]. Multi-scalar theories have also been widely investigated [15] as models of the corrections introduced to general relativity when the standard Einstein-Hilbert action is extended to include gravitational couplings [16], [17] or non-linear curvature terms [18]. Many of these extensions may be conformally transformed [19] to the simple canonical form of single field models, making it important to develop a thorough understanding of the evolution and observable properties of inflationary universes containing single scalar fields.

Previous authors have investigated the consequences when $V(\phi)$ assumes constant [4], polynomial [5], exponential [6] or decaying power-law forms [9]. In a series of recent works we have analysed a wide range of more complicated potentials [12] in an attempt to catalogue the allowed behaviours of these universes and identify features which may be generic to all models. In this article we shall suppose that the universe underwent a phase during which it became dominated by a scalar field possessing a potential of the form,

$$V(\phi) = V_0 \phi^p (\ln \phi)^q \; ; \; V_0, \, p, \, q \text{ constants}. \tag{1.1}$$

Terms such as this appear in the Coleman–Weinberg potential for new inflation [20]. Here we shall study the analytic structure of approximate and exact inflationary universes arising from this family of potentials, and derive their observable properties. We shall focus our attention on the dynamics of the expansion scale factor of the universe $a(t)$, the field $\phi(t)$ and the resulting general properties of the slow-roll phase of the evolution. We will not address issues such as reheating or the exit from inflation in any detail since these features have no bearing on the functional form of the slow-rolling evolution. A full understanding of the forms of slow-roll inflation that can arise from potentials of the form of Eq. (1.1) allows us to complete the catalogue of behaviours that can arise in potentials which incorporate sums and products of powers, exponential, and logarithmic functions of $\phi$.

In section 2 we introduce the equations and variables necessary to describe inflation and the spectra of gravitational wave and density inhomogeneities generated by it. In section 3 we find the different varieties of inflation that can occur for potentials resembling Eq. (1.1) and classify the cosmological behaviours arising for all combinations of the constants $p$ and $q$. In section 4, we give a number of new exact solutions of the Einstein equations which exhibit behaviour related to some of the approximate solutions classified in section 3. In section 5, we derive the spectra and relative intensities of the gravitational wave and density perturbations created by these forms of inflation, and the results are discussed in section 6.

# 2 Einstein's Equations

We study the dynamics of a universe described by the zero curvature Friedmann–Robertson–Walker (FRW) metric with scale factor $a(t)$ and time co-ordinate $t$ in synchronous gauge. We define the Hubble expansion parameter to be $H \equiv \dot{a}/a$, which couples to the material content of the universe through Einstein's equations. For a $k = 0$ FRW universe driven by a classical scalar field $\phi$ these are (in units $8\pi M_{Pl}^{-2} = c = 1$)

$$3H^2 = \frac{1}{2}\dot{\phi}^2 + V(\phi), \tag{2.1}$$

$$\dot{H} = -\frac{1}{2}\dot{\phi}^2, \tag{2.2}$$

$$\ddot{\phi} + 3H\dot{\phi} + V' = 0, \tag{2.3}$$

where $M_{Pl}$ is the Planck mass, primes denote differentiation with respect to the inflaton field $\phi$ and overdots represent derivatives with respect to $t$. $V(\phi)$ is the self-interaction potential characterising the particular form of scalar matter present.

Eqs. (2.1) - (2.3) form a second-order[1], non-linear system in $a$ and $\phi$, which are insoluble in general, given $V(\phi)$. It is possible to make progress by reversing the direction of analysis to postulate a particular $a(t)$ and infer the necessary form for $\phi(t)$, and hence $V(\phi)$, from Einstein's equations [10]. However, the resulting $\phi(t)$ must be invertible; if not, then one cannot deduce $V(\phi)$ analytically. A superior approach is to specify the solution by choosing the functional form of $\phi(t)$ and employing it to deduce $H$ and $V$ from Eqs. (2.1) and (2.2). A good choice will ensure that $\phi(t)$ is invertible and that the first integrals of Eqs. (2.1) and (2.2) may be calculated and so $a(t)$ and $V(\phi)$ can both be found in closed form.

One cannot hope to obtain exact solutions for every conceivable form of scalar field potential $V(\phi)$. In fact, there are very few soluble cases; as a consequence it is usually necessary to resort to the slow-roll approximation [21] which, in its simplest form, requires that the constraints

$$\epsilon_H \equiv 3\frac{\dot{\phi}^2/2}{3H^2} \ll 1, \tag{2.4}$$

$$\eta_H = -3\frac{\ddot{\phi}}{3H\dot{\phi}} \ll 1, \tag{2.5}$$

be satisfied. For reasons that will become apparent, we make a distinction at this stage and refer to these $H$-subscripted quantities as the Hubble Slow-Roll (HSR) parameters. The first of these constraints is identical to the statement $\ddot{a}(t) \gg 0$, and the second guarantees the validity of the first for a prolonged period, justifying the application of this scheme to inflationary dynamics. Applying these conditions to Eqs. (2.1) and (2.3) reduces them to

$$3H^2 \simeq V, \tag{2.6}$$

$$3H\dot{\phi} \simeq -V', \tag{2.7}$$

---
[1] Although we have three first-order equations, any two of them imply the third.

that is,
$$\dot{\phi} \simeq -\frac{V'}{\sqrt{3V}}.\qquad(2.8)$$

The HSR parameters offer a useful tool for understanding the dynamics of inflationary universe models when we possess the relevant exact solution; however, as previously noted, this is seldom true. In these cases we rely upon the Potential Slow-Roll (PSR) constraints to justify our use of Eqs. (2.6)–(2.8). These are

$$\epsilon_V \equiv \frac{1}{2}\left(\frac{V'}{V}\right)^2 \ll 1,\qquad(2.9)$$

$$\eta_V \equiv \frac{V''}{V} \ll 1.\qquad(2.10)$$

To leading order in HSR parameters we have $\epsilon_V = \epsilon_H$ and $\eta_V = \eta_H + \epsilon_H$. The PSR conditions are necessary for a prolonged epoch of inflation, they are not sufficient, however, to ensure its existence. This requires the further assumption that the evolution has reached its limiting inflationary form and we shall assume that $\dot{\phi}$ has reached an attractor solution resembling Eq. (2.8). Finally, we note the PSR parameters for the potential of Eq. (1.1) are

$$\epsilon_V = \frac{1}{2\phi^2}\left\{p + \frac{q}{\ln\phi}\right\}^2,\qquad(2.11)$$

$$\eta_V = \frac{1}{\phi^2}\left\{p(p-1) + \frac{q(2p-1)}{\ln\phi} + \frac{q(q-1)}{(\ln\phi)^2}\right\}.\qquad(2.12)$$

The form of $\epsilon_V$ indicates that inflation will always occur at large $\phi$, but never when $\phi$ is small or close to unity. We shall find that there are solutions in which $\ddot{a} > 0$ and $\dot{a} < 0$ and we refer to these spacetimes which experience decelerated contraction as "conflationary". We also note the class of behaviours which begin (as $t \to -\infty$) in an inflationary state and subsequently cease to inflate ($\epsilon_V < 1$); we shall call these models "deflationary" [22].

## 3 Approximate Solutions

In the absence of a general exact solution we now employ Eqs. (2.6) – (2.8) to map the approximate behaviour of universes containing scalar fields interacting through a potential purely of the form of Eq. (1.1).

For this system Eq. (2.6) is
$$H(\phi) = \sqrt{\frac{V_0}{3}}\phi^{\frac{p}{2}}(\ln\phi)^{\frac{q}{2}},\qquad(3.1)$$

and Eq. (2.8) gives the integral

$$-\sqrt{\frac{V_0}{3}}(t - t_0) = \int_{\phi_o}^{\phi}\frac{\phi^{\frac{2-p}{2}}}{p(\ln\phi)^{\frac{q}{2}} + q(\ln\phi)^{\frac{q}{2}-1}}d\phi.\qquad(3.2)$$

Making the substitution $\psi = \ln \phi$, Eq. (3.2) becomes,

$$-\sqrt{\frac{3}{V_0}}(t - t_0) = \int_{\ln \phi_0}^{\psi} \frac{e^{\frac{4-p}{2}\psi}}{p\psi^{\frac{q}{2}} + q\psi^{\frac{q}{2}-1}} d\psi \equiv I(p, q), \qquad (3.3)$$

where $\phi_0 \equiv \phi(t_0)$. Although we are not able to evaluate $I(p, q)$ for arbitrary $p$ and $q$, we may proceed approximately within discrete regions of the $(p, q)$-parameter space. We first determine the form of $I(p, q)$ at large $\psi$ (i.e. large $\phi$) by means of asymptotic techniques [23].

The qualitative behaviour of the class of potentials defined by Eq. (1.1) and its dependence on the choice of $p$ and $q$ is displayed in Fig. 1. The qualitative structure of the solutions and their corresponding domains of validity in $(p, q)$-parameter space decompose somewhat differently as follows:

## 3.1 The Cases $p \neq 0, 4$

As $\psi \to \infty$ the denominator of the integrand of Eq. (3.3) is governed by the $\psi^{q/2}$ term. We therefore make the approximation

$$I(p, q) \simeq \frac{1}{p} \int_{\ln \phi_0}^{\psi} \frac{e^{\left(\frac{4-p}{2}\right)\psi}}{\psi^{\frac{q}{2}}} d\psi. \qquad (3.4)$$

This category of solutions now subdivides itself further according to sign of $q$. The behaviour when $q = 0$ has been extensively studied [5], [9], here we shall concern ourselves with the instances in which $q$ is non-zero. If $-q/2 \in \mathcal{Z}^+$ then Eq. (3.3) becomes,

$$
\begin{aligned}
t(\phi) &= -\frac{1}{p}\sqrt{\frac{3}{V_0}} \phi^{\frac{4-p}{2}} \left\{ \frac{2}{4-p} (\ln \phi)^{-\frac{q}{2}} \right. \\
&\left. + \sum_{k=1}^{-\frac{q}{2}} \frac{(-1)^k 2^{k+1} \left(-\frac{q}{2}\right)\left(-\frac{q}{2}-1\right)\cdots\left(-\frac{q}{2}-k+1\right)}{(4-p)^{k+1}} (\ln \phi)^{-\frac{q}{2}-k} \right\},
\end{aligned}
\qquad (3.5)
$$

with $t_0$ chosen so as to cancel with the $\phi_0$ arising on the right-hand side. When $q/2 \in \mathcal{Z}^+$ $I(p, q)$ also exists in closed form,

$$
t(\phi) = \sqrt{\frac{3}{V_0}} \left[ \left(\frac{4-p}{2}\right)^{\frac{q-2}{2}} \frac{\mathrm{li}\left(\phi^{\frac{4-p}{2}}\right)}{\left(\frac{q}{2}-1\right)!} \right.
\left. - \phi^{\frac{4-p}{2}} \sum_{k=1}^{\frac{q}{2}-1} \frac{(4-p)^{k-1} (\ln \phi)^{k-\frac{q}{2}}}{2^{k-1} \left(\frac{q}{2}-1\right)\left(\frac{q}{2}-2\right)\cdots\left(\frac{q}{2}-k\right)} \right],
\qquad (3.6)
$$

after a similar fixing of $t_0$; li is the logarithmic integral function (see Appendix 2). To leading order, irrespective of the sign or integer nature of $q/2$, Eq. (3.5) becomes

$$t(\phi) = \frac{2}{p(p-4)} \sqrt{\frac{3}{V_0}} \phi^{\frac{4-p}{2}} (\ln \phi)^{-\frac{q}{2}}, \tag{3.7}$$

as $\phi \to \infty$.

Eq. (3.7) may be inverted asymptotically. To a first approximation we neglect the logarithmic factor to obtain,

$$\phi(t) \simeq \left[ \frac{p(p-4)}{2} \sqrt{\frac{V_o}{3}} t \right]^{\frac{2}{4-p}}. \tag{3.8}$$

Feeding this result back into the logarithm in Eq. (3.7) we obtain the second-order approximation, which at large $\phi$ becomes

$$\phi(t) = (-p)^{\frac{2}{4-p}} \left( \frac{4-p}{2} \right)^{\frac{2-q}{4-p}} \left( \frac{V_0}{3} \right)^{\frac{1}{4-p}} t^{\frac{2}{4-p}} (\ln |t|)^{\frac{q}{4-p}}. \tag{3.9}$$

Knowledge of $\phi(t)$ now enables us to calculate the full time evolution of the system. Using Eq. (3.9) in Eq. (3.1) we have,

$$H(t) = (-p)^{\frac{p}{4-p}} \left( \frac{4-p}{2} \right)^{\frac{p-2q}{4-p}} \left( \frac{V_0}{3} \right)^{\frac{2}{4-p}} t^{\frac{p}{4-p}} (\ln |t|)^{\frac{2q}{4-p}}. \tag{3.10}$$

In the asymptotic limit this may be integrated to obtain the evolution equation for the scale factor,

$$a(t) \propto \exp \left[ \frac{1}{2} (-p)^{\frac{p}{4-p}} \left( \frac{4-p}{2} \right)^{\frac{2q-4}{p-4}} \left( \frac{V_0}{3} \right)^{\frac{2}{4-p}} t^{\frac{4}{4-p}} (\ln |t|)^{\frac{2q}{4-p}} \right]. \tag{3.11}$$

We may treat the scalar field as an effective time variable during the inflationary regime [24], implying a one to one correspondence between $t$ and $\phi$ and enabling us to identify the asymptotic limit in $\phi$ with a unique limit of $t$. The time-solution presented in Eqs. (3.9), (3.10) and (3.11) is valid for $p \neq 0, 4$ as $\phi \to \infty$. The temporal limits here may be obtained by examining the form of Eq. (3.9) and Eq. (3.7) ensures that $t$ has the correct sign to keep the solutions real for all $p$ and $q$. When $p < 0$, $t \to \infty$; when $0 < p < 4$, $t \to -\infty$ and when $p > 4$, $t \to 0$ as $\phi \to \infty$. The last two limits should both be taken to represent the early-time behaviour of the universe. The apparent discrepancy arises because of the different origins of $t$ that we have adopted in each range of $p$ to cancel the integration constants in the solutions for $\phi(t)$. We notice that the partitioning of these regimes is independent of the value of $q$.

## 3.2 The Case $p = 0$

In this class of solutions we retain the latter term in the denominator of Eq. (3.3) and

$$I(p, q) = \frac{1}{q} \int_{\ln \phi_0}^{\psi} \frac{e^{2\psi}}{\psi^{\frac{q-2}{2}}} d\psi. \tag{3.12}$$

This integral may be evaluated for integer values of $1 - q/2$. Once more we encounter a sub-classification of solutions, now dependent upon $q$. When $q \leq 2$ and $1 - q/2 \in \mathcal{Z}^+$, we may perform the integral in Eq. (3.12) to obtain,

$$t(\phi) = \frac{1}{q}\sqrt{\frac{3}{V_0}} \left\{ \frac{2\phi^2}{q-4} \sum_{k=0}^{1-\frac{q}{2}} \frac{(-1)^k \left(2 - \frac{q}{2}\right)\left(1 - \frac{q}{2}\right) \cdots \left(2 - \frac{q}{2} - k\right)}{2^{k+1}} (\ln \phi)^{1-\frac{q}{2}-k} \right\}, \qquad (3.13)$$

choosing $\phi_0$ to cancel $t_0$ on the left-hand side. When $q > 2$ and $q/2 - 1 \in \mathcal{Z}^+$, $I(p,q)$ integrates to yield,

$$t(\phi) = \frac{1}{q}\sqrt{\frac{3}{V_0}} \left\{ \phi^2 \sum_{k=1}^{\frac{q}{2}-2} \frac{2^{k-1} (\ln \phi)^{k-\frac{q}{2}+1}}{\left(\frac{q}{2} - 2\right)\left(\frac{q}{2} - 3\right) \cdots \left(\frac{q}{2} - k\right)} - \frac{2^{\frac{q}{2}-2}}{\left(\frac{q}{2} - 2\right)!} \mathrm{li}\left(\phi^2\right) \right\}. \qquad (3.14)$$

Asymptotically, Eqs. (3.13) and (3.14) converge, independent of the positivity or integer nature of $q/2 - 1$ to yield

$$t(\phi) = -\frac{1}{2q}\sqrt{\frac{3}{V_0}} \phi^2 (\ln \phi)^{1-\frac{q}{2}}. \qquad (3.15)$$

Employing the asymptotic methods used above, we may invert Eq. (3.15) approximately as $\phi \to \infty$ to obtain $\phi(t)$ and hence determine $H(t)$ and $a(t)$. We find,

$$\phi(t) = 2^{\frac{4-q}{4}} \left(\frac{V_0}{3}\right)^{\frac{1}{4}} (-q\, t)^{\frac{1}{2}} (\ln|t|)^{\frac{q-2}{4}} \qquad (3.16)$$

$$H(t) = \sqrt{\frac{V_0}{3}} 2^{-\frac{q}{2}} (\ln|t|)^{\frac{q}{2}} \qquad (3.17)$$

$$a(t) \propto \exp\left\{ \sqrt{\frac{V_0}{3}} 2^{-\frac{q}{2}} t (\ln|t|)^{\frac{q}{2}} \right\}, \qquad (3.18)$$

valid when $p = 0$ as $\phi \to \infty$. Again we may ask at which extremum of $t$ the asymptotic analysis of $\phi$ applies. We find that when $q < 0$, $t \to \infty$ and when $q > 0$, $t \to -\infty$ as $\phi \to \infty$. The choice $q = 0$ leads us back to the well-studied de Sitter universe. When $q = 2$, Eq. (3.16) is an exact solution of Eqs. (2.6) - (2.8) (no asymptotic approximations required). We may calculate the slow-rolling $H(t)$ and $a(t)$, these are

$$H(t) = \frac{1}{2}\sqrt{\frac{V_0}{3}} \ln\left\{ -4\sqrt{\frac{V_0}{3}} t \right\}, \qquad (3.19)$$

$$a(t) \propto \exp\left\{ \frac{1}{2}\sqrt{\frac{V_0}{3}} t \left[ \ln\left(-4\sqrt{\frac{V_0}{3}} t\right) - 1 \right] \right\}. \qquad (3.20)$$

## 3.3 The Case $p = 4$, $q \neq 2$

Here, the integral $I(p, q)$ becomes

$$I(p, q) = \frac{1}{4} \int_{\ln \phi_0}^{\psi} \frac{\psi^{\frac{2-q}{2}}}{\psi + \frac{q}{4}} d\psi. \tag{3.21}$$

If $1 - q/2 \in \mathcal{Z}^{\pm}$ this may be performed exactly. The solutions differ qualitatively with the sign of $1 - q/2$. For $q \neq 2$ we obtain

$$t(\phi) = -\frac{1}{2}\sqrt{\frac{3}{V_0}} \left(\frac{q}{4}\right)^{\frac{3-q}{2}} \left\{ \sum_{k=1}^{1-\frac{q}{2}} \frac{(-1)^{1-\frac{q}{2}+k}}{2k-1} \left(\frac{4}{q}\right)^k (\ln \phi)^{\frac{2k-1}{2}} \right.$$
$$\left. + (-1)^{1-\frac{q}{2}} \tan^{-1}\left[\left(\frac{4}{q}\ln \phi\right)^{\frac{1}{2}}\right] \right\} ; \text{ if } q < 2, \tag{3.22}$$

and

$$t(\phi) = -\frac{1}{2}\sqrt{\frac{3}{V_0}} \left\{ \sum_{k=1}^{\frac{q}{2}-2} \frac{(-1)^k}{q - 2k - 2} \left(\frac{4}{q}\right)^k (\ln \phi)^{k-\frac{q}{2}+1} \right.$$
$$\left. + \frac{(-1)^{\frac{q-4}{2}}}{2} \ln \left| \frac{\ln(\phi)}{\ln(\phi) + \frac{q}{4}} \right| \right\} ; \text{ if } q > 2, \tag{3.23}$$

fixing $t_0$ as usual. In the large $\phi$ limit Eqs. (3.22) and (3.23) assume the simple form,

$$t(\phi) = \frac{1}{2(q-2)} \sqrt{\frac{3}{V_0}} (\ln \phi)^{\frac{2-q}{2}}, \tag{3.24}$$

and by the methods of the previous sections we solve for the full asymptotic time-dependence of the system,

$$\phi(t) = \exp\left\{ [2(q-2)]^{\frac{2}{2-q}} \left(\frac{V_0}{3}\right)^{\frac{1}{2-q}} t^{\frac{2}{2-q}} \right\}, \tag{3.25}$$

$$H(t) = [2(q-2)]^{\frac{2}{2-q}} \left(\frac{V_0}{3}\right)^{\frac{4-q}{2(2-q)}} t^{\frac{q}{2-q}} \exp\left\{ 2^{\frac{4-q}{2-q}}(q-2)^{\frac{2}{2-q}} \left(\frac{V_0}{3}\right)^{\frac{1}{2-q}} t^{\frac{2}{2-q}} \right\}, \tag{3.26}$$

$$a(t) \propto \exp\left\{ 2^{\frac{q}{2-q}}(2-q)^{\frac{4-q}{2-q}} \left(\frac{V_0}{3}\right)^{\frac{4-q}{2(2-q)}} \right.$$
$$\left. \times \exp\left[ 2^{\frac{4-q}{2-q}}(q-2)^{\frac{2}{2-q}} \left(\frac{V_0}{3}\right)^{\frac{1}{2-q}} t^{\frac{2}{2-q}} \right] \right\}, \tag{3.27}$$

valid when $p = 4$ and $q \neq 2$ as $\phi \to \infty$. For the extrema of $t$ we find that as $\phi \to \infty$, $t \to -\infty$ when $q < 2$ and $t \to 0$ when $q > 2$.

## 3.4 The Case $p = 4$, $q = 2$

In this case the form of $I(p, q)$ is particularly simple,

$$I(p, q) = \frac{1}{4} \int_{\ln \phi_0}^{\psi} \frac{d\psi}{\psi + \frac{1}{2}}, \qquad (3.28)$$

leading to

$$t(\phi) = -\frac{1}{4}\sqrt{\frac{3}{V_0}} \ln \left[ \ln \phi + \frac{1}{2} \right]. \qquad (3.29)$$

Without the need to resort to asymptotic analysis, we obtain

$$\phi(t) = \exp \left\{ \exp \left[ -4\sqrt{\frac{V_0}{3}}t \right] - \frac{1}{2} \right\}, \qquad (3.30)$$

$$H(t) = \sqrt{\frac{V_0}{3}} \left[ \exp \left( -4\sqrt{\frac{V_0}{3}}t \right) - \frac{1}{2} \right] \exp \left\{ 2 \exp \left[ -4\sqrt{\frac{V_0}{3}}t \right] - 1 \right\}, \qquad (3.31)$$

$$a(t) \propto \exp \left\{ -\frac{1}{8} \left[ \exp \left( 2e^{-4\sqrt{\frac{V_0}{3}}t} - 1 \right) - e^{-1} \mathrm{E} \left( e^{-4\sqrt{\frac{V_0}{3}}t} \right) \right] \right\}, \qquad (3.32)$$

where E is the exponential integral function, defined in Appendix 2. Eqs. (3.30) – (3.32) form an exact solution to the HSR equations of motion, although we note that in this model $t \to -\infty$ as $\phi \to \infty$ and the asymptotic form of this solution in this region is therefore

$$\phi(t) = \exp \left\{ \exp \left[ -4\sqrt{\frac{V_0}{3}}t \right] \right\}, \qquad (3.33)$$

$$H(t) = \sqrt{\frac{V_0}{3}} \exp \left\{ 2 \exp \left[ -4\sqrt{\frac{V_0}{3}}t \right] \right\}, \qquad (3.34)$$

$$a(t) \propto \exp \left\{ -\frac{1}{8} \exp \left[ 2e^{-4\sqrt{\frac{V_0}{3}}t} \right] \right\}. \qquad (3.35)$$

## 4  Exact Solutions

We seek exact inflationary solutions of the field equations, Eqs. (2.1) – (2.3), in which $\phi(t)$ has the form

$$\phi(t) = A \exp(-\mu t^n). \qquad (4.1)$$

Differentiating with respect to $t$ and eliminating $t$ we obtain,

$$\dot{\phi}(\phi) = (-\mu)^{\frac{1}{n}} n \phi \left[ \ln \left( \frac{\phi}{A} \right) \right]^{\frac{n-1}{n}}. \qquad (4.2)$$

Eq. (2.2) then implies

$$H'(\phi) = \frac{\mu^{\frac{1}{n}} n}{2} \phi \left[ -\ln\left(\frac{\phi}{A}\right) \right]^{\frac{n-1}{n}}, \tag{4.3}$$

leading to,

$$H(\phi) = \frac{\mu^{\frac{1}{n}} n}{2} (-1)^{\frac{n-1}{n}} \int \phi \left[ \ln\left(\frac{\phi}{A}\right) \right]^{\frac{n-1}{n}} d\phi. \tag{4.4}$$

We are now presented with three qualitatively differing classes of behaviour, dependent upon the sign of $(n-1)/n$. From Eq. (4.4) we can see immediately that in models where $1/n$ is not an even integer, exchanging $\mu \leftrightarrow -\mu$ maps between inflationary and conflationary solutions.

## 4.1 The Case $(n-1)/n > 0$ ($n > 1$ or $n < 0$)

When $(n-1)/n$ is an integer (i.e. $n < 0$), we have an exactly integrable system. Eq. (4.4) then gives,

$$\begin{aligned} H(\phi) &= \frac{\mu^{\frac{1}{n}} n}{2} (-1)^{\frac{n-1}{n}} \phi^2 \left(\frac{n}{2n-1}\right) \sum_{k=0}^{\frac{n-1}{n}} \frac{(-1)^k}{2^{k+1}} \left(\frac{n-1}{n} + 1\right)\left(\frac{n-1}{n}\right)\left(\frac{n-1}{n} - 1\right)\ldots \\ &\quad \times \left(\frac{n-1}{n} - k + 1\right) \left[ \ln\left(\frac{\phi}{A}\right) \right]^{1-k-1/n}. \end{aligned} \tag{4.5}$$

In conjunction with Eq. (4.3) this allows calculation of the potential for this theory,

$$\begin{aligned} V(\phi) &= \frac{\mu^{\frac{2}{n}} n^2}{4} \phi^2 \left[ \ln\left(\frac{\phi}{A}\right) \right]^{\frac{2n-2}{n}} \Bigg\{ 3\phi^2 \left(\frac{n}{2n-1}\right)^2 \\ &\quad \times \left[ \sum_{k=0}^{\frac{n-1}{n}} \frac{(-1)^k}{2^{k+1}} \left(\frac{n-1}{n}+1\right)\left(\frac{n-1}{n}\right)\left(\frac{n-1}{n}-1\right)\ldots \right. \\ &\quad \left. \times \left(\frac{n-1}{n} - k + 1\right) \left[\ln\left(\frac{\phi}{A}\right)\right]^{-k} \right]^2 - 2 \Bigg\}, \end{aligned} \tag{4.6}$$

and substituting from Eq. (4.1) we have

$$\begin{aligned} H(t) &= \frac{\mu n}{2} \left(\frac{n}{2n-1}\right) A^2 \exp(-2\mu t^n) \sum_{k=0}^{\frac{n-1}{n}} \frac{(-1)^k}{2^{k+1}} \\ &\quad \times \left(\frac{n-1}{n}+1\right)\left(\frac{n-1}{n}\right)\left(\frac{n-1}{n}-1\right)\ldots \\ &\quad \times \left(\frac{n-1}{n} - k + 1\right) [-\mu t^n]^{-k}. \end{aligned} \tag{4.7}$$

Making the substitution $\theta = t^n$, we can integrate Eq. (4.7) to obtain the scale factor evolution,

$$a(t) \propto \exp\left\{\frac{\mu n A^2}{2(2n-1)}\left[\sum_{k=1}^{\frac{n-1}{n}}\left(\frac{1}{\mu}\right)\frac{1}{2^{k+1}}\left(\frac{n-1}{n}+1\right)\left(\frac{n-1}{n}\right)\left(\frac{n-1}{n}-1\right)\cdots\right.\right.$$
$$\times\left(\frac{n-1}{n}-k+1\right)\left[\frac{(-2\mu)^{k-1}}{(k-1)!}E(-2\mu t^n) - \exp(-2\mu t^n)\right.$$
$$\left.\times\sum_{r=1}^{k-1}\frac{(-2\mu)^{r-1}}{(k-1)(k-2)\ldots(k-r)t^{n(k-r)}}\right]$$
$$\left.\left.+\frac{1}{2\mu}\left(\frac{1-2n}{2n}\right)\exp(-2\mu t^n)\right]\right\}. \qquad (4.8)$$

We illustrate this class of solutions by considering the relatively simple case $n = -1$, for which $(n-1)/n = 2$. We find the $\phi$-parameterised solution,

$$H(\phi) = -\frac{\phi^2}{4\mu}\left\{\left[\ln\left(\frac{\phi}{A}\right)\right]^2 - \ln\left(\frac{\phi}{A}\right) + \frac{1}{2}\right\}, \qquad (4.9)$$

$$V(\phi) = \frac{3\phi^4}{16\mu^2}\left\{\left[\ln\left(\frac{\phi}{A}\right)\right]^2 - \ln\left(\frac{\phi}{A}\right) + \frac{1}{2}\right\}^2 - \frac{\phi^2}{2\mu^2}\left[\ln\left(\frac{\phi}{A}\right)\right]^4, \qquad (4.10)$$

and, in terms of $t$, this is

$$H(t) = -\frac{A^2}{4\mu}\left\{\left(\frac{\mu}{t}\right)^2 + \frac{\mu}{t} + \frac{1}{2}\right\}\exp\left(-\frac{2\mu}{t}\right), \qquad (4.11)$$

$$a(t) = \exp\left\{-\frac{A^2}{8}\left[1+\frac{t}{\mu}\right]\exp\left(-\frac{2\mu}{t}\right)\right\}. \qquad (4.12)$$

The potential in Eq. (4.10), and $\epsilon_H$ as derived from Eqs. (2.4) and (4.11) are displayed in Fig. 2. To ensure $V(\phi) > 0$ as required for inflation, we have confined our interest to the domain $\phi > 0$. Although $V(\phi)$ and $\epsilon_H$ are independent of the sign of $\mu$ this is not the case for the remaining dynamical quantities. When $\mu < 0$ the field rolls from $\phi = \infty$ at $t = 0$ to $\phi = A$ as $t \to \infty$ and from either of Eqs. (4.9) and (4.11) we see that $H > 0$ for all $\phi$ and $t$ values and the universe is expanding. Examination of the form of $\epsilon_H$ reveals that when $A \lesssim 2.4$ the evolution is intially inflationary, temporarily deflating as the universe becomes kinetic-dominated, before re-inflating in the final stages of the evolution as $\phi \to A$. This double-inflation feature is not generic; when $A > 2.4$, $\epsilon_H < 1$ for all $\phi$, $t$. The scalar curvature, $\mathcal{R} = 12(H'^2 - H^2)$, is singular at $t = 0$ in this model and approaches the constant $\mathcal{R}_* = -3A^4/16\mu^2$ at late times. It also contains a maximum, the height and position of which are determined by the values of $A$ and $\mu$.

When $\mu > 0$ the field starts at $\phi = 0$ and rolls asymptotically to $\phi = A$. $H$ is always negative in this case and the system describes the dynamics of a collapsing universe. The HSR parameter $\epsilon_H$ is invariant of the sign of $\mu$ and so we have the possibility of two conflationary epochs when $A \lesssim 2.4$. The Ricci curvature $\mathcal{R}$ in this case is initially zero, passes through a maximum and tends to $\mathcal{R}_*$ as $t$ becomes large.

As a final observation, regarding this case we note the approach of the potential to the form of Eq. (1.1) at large $\phi$ since $V(\phi) \to 3\phi^4(\ln\phi)^4/16\mu^2$ as $\phi \to \infty$, and its tendency toward a simple polynomial form as $\phi \to A$.

## 4.2 The Case $(n-1)/n < 0$ $(0 < n < 1)$

Once again, the solution is only integratable if $(n-1)/n$ is an integer, i.e. $0 < n < 1$; we now explore this possibility. Eq. (4.4) integrates to give

$$H(\phi) = \frac{\mu^{\frac{1}{n}} n}{2} (-1)^{\frac{n-1}{n}} A^2 \left[ \frac{2^{\frac{1-2n}{n}}}{\left(\frac{1-2n}{n}\right)!} \mathrm{li}\left[\left(\frac{\phi}{A}\right)^2\right] \right.$$

$$\left. - \left(\frac{\phi}{A}\right)^2 \sum_{k=1}^{\frac{1-2n}{n}} \frac{2^{k-1}}{\left(\frac{1-n}{n} - 1\right)\left(\frac{1-n}{n} - 2\right)\ldots\left(\frac{1-n}{n} - k\right)} \right.$$

$$\left. \times \left[\ln\left(\frac{\phi}{A}\right)\right]^{\frac{n-1}{n}+k} \right]. \tag{4.13}$$

The potential driving the evolution in this case is

$$V(\phi) = \frac{\mu^{\frac{2}{n}} n^2}{4} \left[ 3A^4 \left\{ \frac{2^{\frac{1-2n}{n}}}{\left(\frac{1-2n}{n}\right)!} \left[\mathrm{li}\left(\frac{\phi}{A}\right)^2\right] \right.\right.$$

$$\left.\left. - \left(\frac{\phi}{A}\right)^2 \sum_{k=1}^{\frac{1-2n}{n}} \frac{2^{k-1}}{\left(\frac{1-n}{n} - 1\right)\left(\frac{1-n}{n} - 2\right)\ldots\left(\frac{1-n}{n} - k\right)} \right.\right.$$

$$\left.\left. \times \left[\ln\left(\frac{\phi}{A}\right)\right]^{\frac{n-1}{n}+k} \right\}^2 - 2\phi^2 \left[\ln\left(\frac{\phi}{A}\right)\right]^{\frac{2n-2}{n}} \right]. \tag{4.14}$$

The $\phi$-dependence in Eq. (4.13) may be converted into a time dependence using Eq. (4.1),

$$H(t) = \frac{\mu n}{2} A^2 \left[ \frac{2^{\frac{1-2n}{n}}}{\left(\frac{1-2n}{n}\right)!} \mathrm{E}\left(-2\mu t^n\right) \right.$$

$$\left. - \exp\left(-2\mu t^n\right) \sum_{k=1}^{\frac{1-2n}{n}} \frac{2^{k-1}}{\left(\frac{1-n}{n} - 1\right)\left(\frac{1-n}{n} - 2\right)\ldots\left(\frac{1-n}{n} - k\right)} \right.$$

$$\times \ (-\mu t^n)^k \Bigg] \ . \tag{4.15}$$

We can integrate to obtain the behaviour of the scale factor,

$$a(t) \ \propto \ \exp\Bigg\{\frac{\mu^{\frac{1}{n}} n}{2}(-1)^{\frac{n-1}{n}} A^2 \Bigg[\frac{2^{\frac{1-2n}{n}}}{\left(\frac{1-2n}{n}\right)!} \Bigg\{ t\,\mathrm{E}\left(-2\mu t^n\right) + \exp\left(-2\mu t^n\right)$$

$$\times \Bigg[\frac{t^{1-n}}{2\mu} - \sum_{k=1}^{\frac{1-n}{n}} \frac{(-1)^k}{(-2\mu)^{k+1}}\left(\frac{1-n}{n}\right)\left(\frac{1-n}{n}-1\right)\ldots\left(\frac{1-n}{n}-k+1\right)$$

$$\times t^{1-n-nk}\Bigg]\Bigg\} - \sum_{k=1}^{\frac{1-2n}{n}} \frac{2^{k-1}(-\mu)^{\frac{n-1}{n}+k}}{\left(\frac{1-n}{n}-1\right)\left(\frac{1-n}{n}-2\right)\ldots\left(\frac{1-n}{n}-k\right)} \frac{\exp\left(-2\mu t^n\right)}{n}$$

$$\times \Bigg[\sum_{r=1}^{k} \frac{(-1)^r}{(-2\mu)^{r+1}} k(k-1)\ldots(k-r+1)\, t^{n(k-r)} - \frac{t^{kn}}{2\mu}\Bigg] \Bigg\} \ . \tag{4.16}$$

As a simple example of these models we study the special case $n = 1/3$ i.e., $(n-1)/n = -2$. We find,

$$H(\phi) \ = \ \frac{\mu^3 A^2}{6}\left\{2\,\mathrm{li}\left[\left(\frac{\phi}{A}\right)^2\right] - \frac{\phi^2}{A^2 \ln\left(\frac{\phi}{A}\right)}\right\} \ , \tag{4.17}$$

$$V(\phi) \ = \ \frac{\mu^6 A^4}{12}\left\{2\,\mathrm{li}\left[\left(\frac{\phi}{A}\right)^2\right] - \frac{\phi^2}{A^2 \ln\left(\frac{\phi}{A}\right)}\right\}^2 - \frac{\mu^6 \phi^2}{18\left[\ln\left(\frac{\phi}{A}\right)\right]^4} \ , \tag{4.18}$$

and,

$$H(t) \ = \ \frac{A^2 \mu^3}{6}\left\{\frac{\exp\left(-2\mu t^{1/3}\right)}{\mu t^{1/3}} + 2\mathrm{E}\left(-2\mu t^{1/3}\right)\right\} \ , \tag{4.19}$$

$$a(t) \ \propto \ \exp\Bigg\{\frac{A^2 \mu^3}{3} t\mathrm{E}\left(-2\mu t^{1/3}\right) + \frac{A^2 \mu^2}{6} t^{2/3} \exp\left(-2\mu t^{1/3}\right) - \frac{A^2 \mu}{12} t^{1/3} \exp\left(-2\mu t^{1/3}\right)$$

$$-\frac{A^2}{24}\exp\left(-2\mu t^{1/3}\right)\Bigg\} \ . \tag{4.20}$$

Although these models are all inflationary in the limit as $\phi \to \infty$, the potentials driving them only resemble the form of Eq. (1.1) at small $\phi$ and it can be seen from the plot of $\epsilon_H$ for these theories in Fig. 3 that the evolution is non-inflationary in this limit. Despite this, the interim behaviour between $\phi = A$ and $\phi = 0$ has a particularly interesting structure.

From Fig. 3 we see that $\epsilon_H$ contains a pronounced minimum when $0 < \phi < A$ and one can show that when $A \gtrsim 11.55$, $\epsilon_H(\phi_{\mathrm{min}}) \lesssim 1$; indicative of accelerated behaviour in the trough of the minimum. If we confine our attention to models in which $\mu > 0$, then $H > 0$, and the universe expands from a curvature singularity as the field rolls from $\phi = A$ at $t = 0$ to $\phi = 0$ as $t \to \infty$. This model encompasses both the beginning and the end of the inflationary regime and is exact. Inflation here is a transient phenomenon, arising naturally as $\dot{\phi}$ decays from its initially kinetic-dominated state and ending once $\dot{\phi}$ becomes large again.

The form of $V(\phi)$ in this case is reminiscent of the potential for intermediate inflation [9], as may be seen from Fig. 3. As the field approaches $\phi = 0$ the potential approaches zero much faster than in any intermediate inflationary scenario and this serves to halt inflation after finite duration.

### 4.3 The Case $(n-1)/n = 0$ $(n = 1)$

In this case we obtain the simple solution,

$$H(\phi) = \frac{\mu}{4}\phi^2, \tag{4.21}$$

$$V(\phi) = \frac{\mu^2}{2}\phi^2\left(\frac{3}{8}\phi^2 - 1\right), \tag{4.22}$$

$$H(t) = \frac{\mu A^2}{4}\exp(-2\mu t), \tag{4.23}$$

$$a(t) \propto \exp\left[-\frac{A^2}{8}e^{-2\mu t}\right]. \tag{4.24}$$

When $\mu$ is positive, the universe expands from a curvature singularity at $t = -\infty$, displays slow-rolling inflation until a time $t_* = \frac{1}{2\mu}\ln\frac{A^2}{8}$, and then tends to the static Minkowski form as $t$ becomes large. This is an example of a deflationary universe, asymptoting to a Friedmann model with a constant scale factor. For negative $\mu$ this scenario runs in reverse, collapsing from maximum extent at $t = -\infty$ to become conflationary at $t_*$, before ending in a "big crunch" as $t \to \infty$.

## 5 Observable Properties

Inflation offers an attractive mechanism to generate perturbations that may have seeded the formation of galaxies and large scale structure observed in the present day universe.

The current theory of the magnification of quantum mechanical fluctuations and their subsequent growth upon horizon re-entry via gravitational instability has been extensively examined [2]. We quantify the scale dependence of scalar and tensor metric perturbations in terms of the spectral indices $n_s$ and $n_g$ respectively. During slow-roll inflation we may express these quantities as expansions in terms of the PSR parameters and to first-order we obtain [2],

$$1 - n_s = 6\epsilon_V - 2\eta_V, \tag{5.1}$$

$$n_g = -2\epsilon_V, \tag{5.2}$$

$$R_l = \frac{25}{2}\epsilon_V. \tag{5.3}$$

The second-order corrections [25] to these results are cumbersome and not particularly illuminating and so we will not introduce them here. Furthermore, one may show that the extension to the domain of validity of the results offered by the second-order corrections is small, justifying a first-order analysis. Since the PSR parameters are deemed to be small during inflation, the generic predictions should be $n_s \sim 1$, $n_g \sim R_l \sim 0$. For the potential described by Eq. (1.1), Eqs. (5.1) - (5.3) become

$$1 - n_s = \frac{1}{\phi^2}\left\{p(p+2) + \frac{2q(p+1)}{\ln\phi} + \frac{q(q+2)}{(\ln\phi)^2}\right\}, \tag{5.4}$$

$$n_g = -\frac{1}{\phi^2}\left(p + \frac{q}{\ln\phi}\right)^2, \tag{5.5}$$

$$R_l = \frac{25}{4\phi^2}\left(p + \frac{q}{\ln\phi}\right)^2, \tag{5.6}$$

and there are obvious simplifications when $p = -2, -1, 0$ and $q = 0, -2$. The form of these expressions confirm that we always observe slow-rolling inflation at large $\phi$ and the values of the observables become asymptotically indistinguishable from the pure de Sitter case in this limit. When $\phi$ is smaller, however, it becomes possible for $n_s$ to deviate from unity, and for the gravitational wave contribution to the perturbation spectrum to become large. We also note that when $-2 < p < 0$, $n_s$ can rise above unity, introducing additional power at small scales in a manner akin to intermediate inflation [9]. This can occur whenever $\eta_H > 3\epsilon_H$ and arises as a result of slow-roll evolution that is not friction dominated, i.e. when $\dot{\phi}^2$ is small, but $|\ddot{\phi}|$ is allowed to vary. In these cases we note the restriction $n_s \lesssim 1.5$, based on estimates of the overproduction of primordial black holes in such a scenario [26].

The models presented here make no provision for a natural end to the inflationary epoch, and for this reason it is not possible to invoke constraints on $p$ and $q$ from the requirement of minimum expansion, or that the reheating temperature be large enough to allow baryogenesis to occur [1]. Despite this, bounds may be placed on the energy scale, $V_0$, at which inflation occurs [27], firstly by enforcing the consistency relation for our classical analysis, $3H^2(\phi) \lesssim M_{Pl}^4$, and more importantly by demanding that the predicted $\delta\rho/\rho$ be consistent with the CMBR quadrupole anisotropy, $\delta\rho/\rho \simeq 2.3 \times 10^{-5}$, as measured by the COsmic Background Explorer (COBE) satellite [3].

# 6  Conclusions

In this article we have studied a number of exact and approximate inflationary universes, obtained as solutions to Einstein's equations when one postulates a form for the inflaton potential resembling Eq. (1.1). We have derived the functional form of the spectral indices arising in this case and have indicated how further constraints can be imposed by comparison with observations. We show that when the potential also contains terms of the form li($\phi^2$) one may obtain exact solutions interpolating smoothly the epochs of initial kinetic-dominated expansion, potential-dominated inflation and late-time Friedmann like expansion. This extends the deflationary universe models of [22] to include cases in which the initial evolution is non-inflationary. Generally, however, the solutions are not complete in this way and are only meant to describe evolution on the finite portion of the real potential where inflation occurs. The mechanism by which the universe exits the inflationary epoch and enters the reheating phase is not treated here, instead we have sought to classify the possible ways in which an inflationary universe containing a slow-rolling scalar field with a very general potential may behave. Our results complete a library of solutions for the evolution of slow-rolling scalar field universes when the potential is a combination of logarithms, powers and exponentials of the field $\phi$ [12], [14] and are summarised in Table 1. Other variants, consistent with slow-rolling, may be obtained by using these results in conjunction with the methods described in [14] to effect transformations of known inflationary solutions which introduce new functional dependences while preserving the slow-roll nature of the solution. The final entry in Table 1 was obtained in this manner and the full solution is presented in Appendix 1.

| $V(\phi)$ | Range | Inflation | Scale factor |
|---|---|---|---|
| $V_0 \phi^n \exp(-\lambda \phi^m)$ | $m = n = 0$ | always | $\exp\left[\sqrt{\frac{V_0}{3}}t\right]$ |
| | $m = 1,\ n = 0$ | if $\lambda^2 < 2$ | $t^{2/\lambda^2}$ |
| | $0 < m < 1$ | as $t \to \infty$ | $\exp\left[\frac{1}{\lambda m(2-m)}\left(\frac{2}{\lambda}\right)^{\frac{2-m}{m}} \ln^{\frac{2-m}{m}} t\right]$ |
| | $m \le 0,\ n \ne 0,\ 4$ | $n > 4,\ t \to 0$ | $\exp\left[\left(\frac{V_0}{3}\right)^{\frac{2}{4-n}}(-2n)^{\frac{n}{4-n}}\left(\frac{4-n}{4}\right)^{\frac{4}{4-n}}\right.$ |
| | | $0 < n < 4,\ t \to -\infty$ | $\left. \times t^{\frac{4}{4-n}}\right]$ |
| | | $n < 0,\ t \to \infty$ | |
| | $m \le 0,\ n = 0$ | as $t \to -\infty$ | $\exp\left\{\sqrt{\frac{V_0}{3}}t\exp\left[-\frac{\lambda}{2}\right.\right.$ |
| | $m > 1,\ n = 0$ | $1 < m < 2,\ t \to 0$ | $\left.\left.\times\left\{\lambda m(2-m)\sqrt{\frac{V_0}{3}}\right\}^{\frac{m}{2-m}} t^{\frac{m}{2-m}}\right]\right\}$ |
| | | $m > 2,\ t \to -\infty$ | |
| | $m = 2,\ n = 0$ | as $t \to -\infty$ | $\exp\left\{\sqrt{\frac{V_0}{3}}t\exp\left[-\frac{\lambda}{2}\exp\left(4\sqrt{\frac{V_0}{3}}\lambda t\right)\right]\right\}$ |
| | $m \le 0,\ n = 4$ | as $t \to -\infty$ | $\exp\left[-\frac{1}{8}\exp\left\{-8\sqrt{\frac{V_0}{3}}t\right\}\right]$ |
| $V_0 \phi^p (\ln \phi)^q$ | $p \ne 0,\ 4$ | $p < 0,\ t \to \infty$ | $\exp\left[\frac{1}{2}(-p)^{\frac{p}{4-p}}\left(\frac{4-p}{2}\right)^{\frac{2q-4}{p-4}}\left(\frac{V_0}{3}\right)^{\frac{2}{4-p}}\right.$ |
| | | $0 < p < 4,\ t \to -\infty$ | $\left.\times t^{\frac{4}{4-p}}(\ln|t|)^{\frac{2q}{4-p}}\right]$ |
| | | $p > 4,\ t \to 0$ | |
| | $p = 0$ | $q < 0,\ t \to \infty$ | $\exp\left\{\sqrt{\frac{V_0}{3}}2^{-\frac{q}{2}}t(\ln|t|)^{\frac{q}{2}}\right\}$ |
| | | $q > 0,\ t \to -\infty$ | |
| | $p = 4,\ q \ne 2$ | $q < 2,\ t \to -\infty$ | $\exp\left\{2^{\frac{q}{2-q}}(2-q)^{\frac{4-q}{2-q}}\left(\frac{V_0}{3}\right)^{\frac{4-q}{2(2-q)}}\right.$ |
| | | $q > 2,\ t \to 0$ | $\left.\times \exp\left[2^{\frac{4-q}{2-q}}(q-2)^{\frac{2}{2-q}}\left(\frac{V_0}{3}\right)^{\frac{1}{2-q}} t^{\frac{2}{2-q}}\right]\right\}$ |
| | $p = 4,\ q = 2$ | as $t \to -\infty$ | $\exp\left\{-\frac{1}{8}\exp\left[2e^{-4\sqrt{\frac{V_0}{3}}t}\right]\right\}$ |
| $V_0 (\ln \phi)^r e^{-\lambda \phi}$ | | $\lambda^2 < 2,\ t \to \infty$ | $t^{2/\lambda^2}[\ln(\ln t)]^{q/\lambda^2}$ |

Table 1: A summary of the evolution of the scale factor of inflationary universes driven by scalar fields with potential $V(\phi)$.

# Acknowledgements

The authors were supported by the PPARC. PP would like to thank Andrew Laycock for discussions.

# Appendix 1 - Logarithmic-Exponential Potentials

Here we briefly describe the behaviour when an exponential potential is modified by the inclusion of a logarithmic factor. This possibility was first explored in [14]. We find that when $V(\phi) = V_0 (\ln \phi)^q \exp(-\lambda \phi)$, the time-dependent solution, at large $\phi$ and $t$, is

$$\phi(t) = \frac{2}{\lambda} \ln t + \frac{2}{\lambda} \ln \left[ \ln^{1/2q} \{\ln t\} \right], \tag{A1.1}$$

$$H(t) = \frac{2}{\lambda^2 t} \left[ 1 + \left[ \frac{\ln \left\{ \ln \left[ \ln^{1/2q} (\ln t) \right] \right\}}{\ln(\ln t)} \right]^{q/2} \right], \tag{A1.2}$$

$$a(t) \propto t^{2/\lambda^2} \ln^{q/\lambda^2}(\ln t). \tag{A1.3}$$

Variants of this potential are displayed in Fig. 4.

# Appendix 2 - Special Functions and Approximations

We define the exponential integral function $\mathrm{E}(a\phi)$ to be

$$\mathrm{E}(a\phi) \equiv \mathrm{E}_1(a\phi) \equiv \int_{\phi}^{\infty} \frac{e^{-a\tau}}{\tau} d\tau \quad \text{for} \quad -\infty < a\phi < 0 \tag{A2.1}$$

$$\equiv \mathrm{Ei}(a\phi) \equiv \int_{-\infty}^{\phi} \frac{e^{a\tau}}{\tau} d\tau \quad \text{for} \quad 0 < a\phi < \infty, \tag{A2.2}$$

and $\mathrm{E}(a\phi)$ is singular at $\phi = 0$. Similarly the logarithmic integral, $\mathrm{li}(a\phi)$, is defined as

$$\mathrm{li}(a\phi) \equiv a \int_{0}^{\phi} \frac{d\tau}{\ln(a\tau)} \quad \text{for} \quad a\phi \geq 0. \tag{A2.3}$$

Eqs. (A2.1), (A2.2) and (A2.3) are linked by the identity

$$\mathrm{li}(a\phi) \equiv \mathrm{E}[\ln(a\phi)]. \tag{A2.4}$$

We note the approximate relations,

$$\mathrm{E}(a\phi) \to \frac{e^{a\phi}}{a\phi} \quad \text{as} \quad \phi \to -\infty, \tag{A2.5}$$

$$\mathrm{li}(a\phi) \to \frac{a\phi}{\ln \phi} \quad \text{as} \quad \phi \to 0, \tag{A2.6}$$

and the rather more crude asymptotic formulae, asserting the approximate validity of Eqs. (A2.5) and (A2.6) as $\phi \to \infty$.

# Figure Captions

Figure 1: The potential $V(\phi) = V_0 \phi^p (\ln \phi)^q$ when $V_0 = 1$ for the parameter choices: $p=2$ (a. and b.), $p=-2$ (c. and d.), $q=1, 2, 3$ (a. and c.) and $q=-1, -2, -3$ (b. and d.). The quantities $\phi$ and $V(\phi)$ are in Planck units.

Figure 2: a. $V(\phi)$ and b. $\epsilon_H$, for the parameter choices $A = 1.5$, $n = -1$ (all axes in Planck units).

Figure 3: a. $V(\phi)$ and b. $\epsilon_H$ in the range $0 < \phi < A$ for the parameter choices $|\mu| = 1$, $A = 12$ and $n = 1/3$ (all axes in Planck units).

Figure 4: The potential $V(\phi) = V_0 (\ln \phi)^q \exp(-\lambda \phi)$ when $V_0 = \lambda = 1$ for: a. positive $q$ and b. negative $q$ (all axes in Planck units).

# References

[1] E. W. Kolb and M. S. Turner, *The Early Universe*, Addison-Wesley, (1990).
A. D. Linde, *Particle Physics and Inflationary Cosmology*, Gordon and Breach, (1990).

[2] A. R. Liddle and D. H. Lyth, Phys. Rep. **231**, 1 (1993).

[3] G. F. Smoot *et al*, Astrophys. J. Lett. **396**, L1 (1992).
K. Górski *et al*, Astrophys. J. Lett. **430**, L89 (1994).

[4] A. H. Guth, Phys. Rev. D**23**, 347 (1981).

[5] A. D. Linde, Phys. Lett. B**129**, 177 (1983).

[6] L. F. Abbott and M. B. Wise, Nucl. Phys. **B244**, 541 (1984).
J. D. Barrow, A. B. Burd and D. Lancaster, Class. Quantum Grav. **3**, 551 (1985).
F. Lucchin and S. Matarrese, Phys. Rev. D**32**, 1316 (1985).
J. J. Halliwell, Phys. Lett. B**185**, 341 (1987).
J. D. Barrow, *ibid.* **187**, 341 (1987).
A. B. Burd and J. D. Barrow, Nucl. Phys. **B308**, 929 (1988).

[7] D. S. Salopek and J. R. Bond, Phys. Rev. D**42**, 3936 (1990).

[8] A. G. Muslimov, Class. Quant. Grav. **7**, 231 (1990).

[9] J. D. Barrow, Phys. Lett. B**235**, 40 (1990).
J. D. Barrow and P. Saich, Phys. Lett. B**249**, 406 (1990).
J. D. Barrow and A. R. Liddle, Phys. Rev. D**47**, R5129 (1993).

[10] G. F. R. Ellis and M. S. Madsen, Class. Quantum Grav. **8**, 667, (1991).

[11] J. D. Barrow and P. Saich, Class. Quantum Grav. **10**, 279 (1993).

[12] J. D. Barrow, Phys. Rev. D**48**, 1585 (1993).
J. D. Barrow, Phys. Rev. D**49**, 3055 (1994).
P. Parsons and J. D. Barrow, "Generalized Scalar Field Potentials and Inflation", SUSSEX-AST 95/1-1, astro-ph/9501086, to appear Phys. Rev. D June 1995.

[13] F. E. Schunck and E. W. Mielke, Phys. Rev. D**50**, 4794 (1994).

[14] P. Parsons and J. D. Barrow, "New Inflation From Old", Sussex Preprint (1995), to appear Class. Quantum Grav..

[15] A. D. Linde, Phys. Rev. D **49**, 748 (1994).
A. L. Berkin and R. W. Hellings, *ibid.* **49**, 6442 (1994).
E. J. Copeland, A. R. Liddle, D. H. Lyth, E. D. Stewart and D. Wands, *ibid.* **49**, 6410 (1994).
D. Roberts, A. R. Liddle and D. H. Lyth, *ibid.* **51**, 4122 (1995).

[16] J. D. Barrow and K. Maeda, Nucl. Phys. **B341**, 294 (1990).
J. D. Barrow and J. Mimoso, Phys. Rev. D **50**, 3746 (1994).
J. D. Barrow, *ibid.* **51**, 2729 (1995).

[17] A. M. Laycock and A. R. Liddle, Phys. Rev. D **49**, 1827 (1994).
A. M. Laycock, "The Kinematics and Dynamics of Non-Minimally Coupled Scalar Fields", Sussex Preprint (1995).

[18] J. D. Barrow and A. Ottewill, J. Phys. **A16**, 2757 (1983).
A. A. Starobinsky, Sov. Astron. Lett. **9**, 302 (1983).

[19] J. D. Barrow and S. Cotsakis, Phys. Lett. **B214**, 515 (1988).
K. Maeda, Phys. Rev. D **37**, 858 (1988).

[20] A. D. Linde, Phys. Lett. **108B**, 389 (1982).
A. Albrecht and P. J. Steinhardt, Phys. Rev. Lett. **48**, 1220 (1982).

[21] P. J. Steinhardt and M. S. Turner, Phys. Rev. D**29**, 2162 (1984).
A. R. Liddle and D. H. Lyth, Phys. Lett. **B291**, 391 (1992).
A. R. Liddle, P. Parsons and J. D. Barrow, Phys. Rev. D**50**, 7222 (1994).

[22] J. D. Barrow, Phys. Lett. **B180**, 335 (1986).
J. D. Barrow, *ibid.* **183**, 285 (1987).
J. D. Barrow, Nucl. Phys. **B310**, 743 (1988).
N. Turok, Phys. Rev. Lett. **60**, 549 (1988).
J. D. Barrow, in *The Formation and Evolution of Cosmic Strings*, ed. G. W. Gibbons, S. W. Hawking and T. Vachaspati, CUP (1990).
D. Pavón, J. Bafaluy and D. Jou, Class. Quantum Grav. **8**, 347 (1991).

[23] F. W. J. Olver, *Asymptotics and Special Functions*, Academic Press, (1974).

[24] J. E. Lidsey, Phys. Lett. **B273**, 42 (1991).

[25] E. D. Stewart and D. H. Lyth, Phys. Lett. **B302**, 171 (1993).

[26] B. J. Carr and J. E. Lidsey, Phys. Rev. **D48**, 543 (1993).
B. J. Carr, J. H. Gilbert and J. E. Lidsey, *ibid.* **50**, 4853 (1994).

[27] A. R. Liddle, Phys. Rev. **D49**, 739 (1994).

[28] *Handbook of Mathematical Functions*, ed. M. Abramowitz and I. A. Stegun, Natl. Bur. Stand. Appl. Math. Ser. No. 55 (U. S. GPO, washington, D. C., 1965).

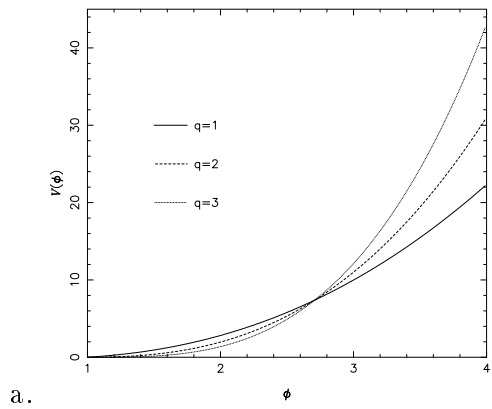
a.

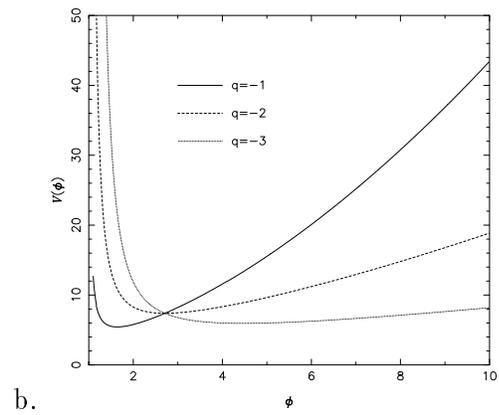
b.

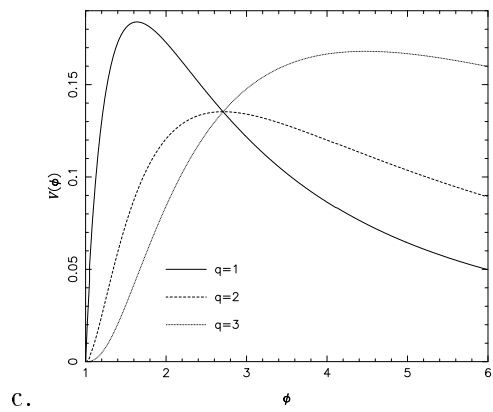
c.

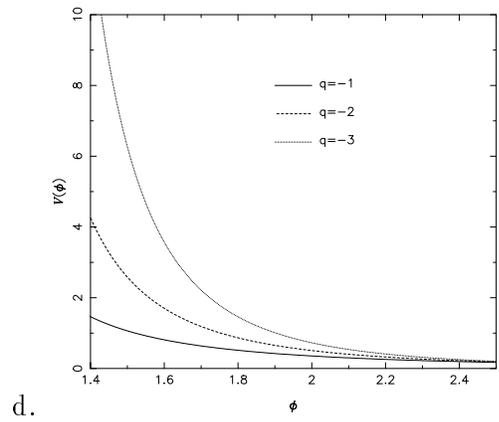
d.

Figure 1

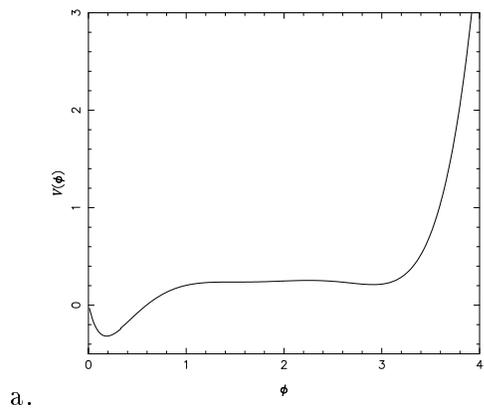 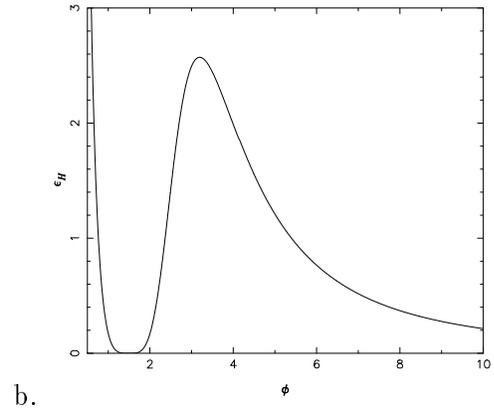

a. b.

Figure 2

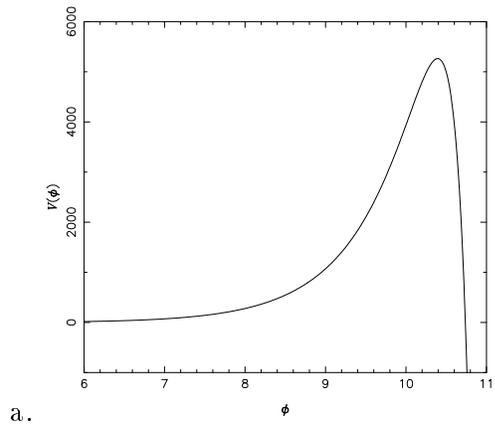
a.
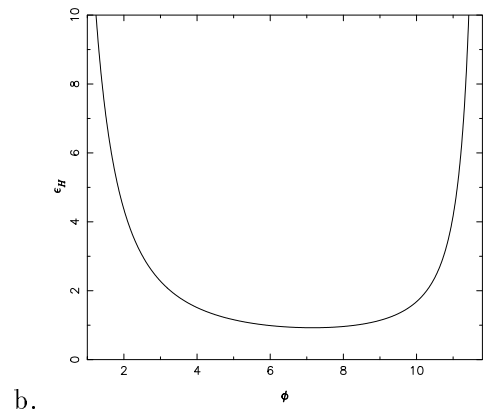
b.

Figure 3

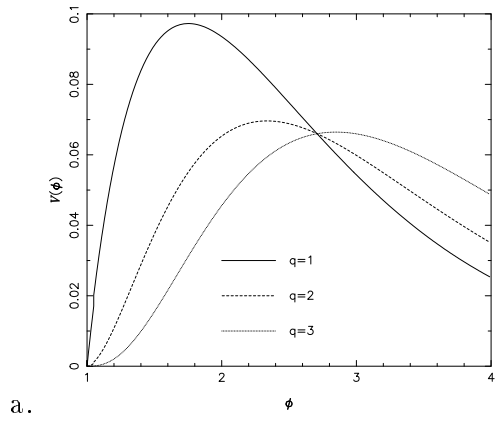 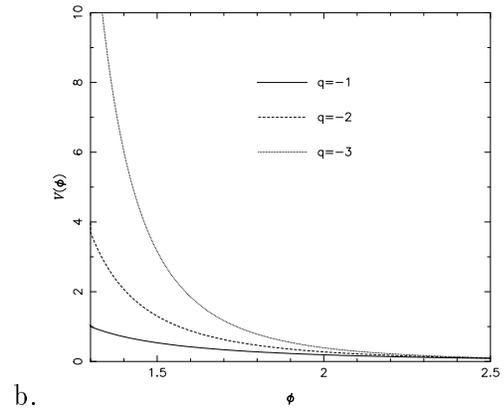

a.  b.

Figure 4